\documentclass[prb,aps,twocolumn,epsfig,eqsecnum]{revtex4}
\usepackage{epsfig}
\usepackage{wasysym}
\usepackage{bm}
\usepackage{pstricks}
\newlength{\myL}

\newcommand{\hll}{pentagonal lattice}

\newcommand{\beq}{\begin{equation}}
\newcommand{\eeq}{\end{equation}}
\newcommand{\bea}{\begin{eqnarray}}
\newcommand{\eea}{\end{eqnarray}}

\newcommand{\etal}{{\em et al.}}

\def\tit#1#2#3#4#5{{#1}{\bf #2}, #3 (#4)}
\def\jmp{J.\ Math.\ Phys.\ }

\def\prl{Phys.\ Rev.\ Lett.\ }
\def\pr{Phys.\ Rev.\ }
\def\prb{Phys.\ Rev.\ B\ }

\def\jpco{J.\ Phys.\ Cond.\ Mat.\ }

\def\sci{Science\ }

\begin{document}

\title{SU(2)-invariant 
spin-1/2 Hamiltonians with RVB and other valence bond phases}

\author{K. S. Raman$^1$, R. Moessner$^2$ and S. L. Sondhi$^1$}

\affiliation{$^1$Department of Physics, Princeton University,
Princeton, NJ 08544, USA}

\affiliation{$^2$Laboratoire de Physique Th\'eorique de l'Ecole Normale
Sup\'erieure, CNRS-UMR8549, Paris, France}

\date{\today}

\begin{abstract}

We construct a family of rotationally invariant, local, S=1/2
Klein Hamiltonians on various lattices that exhibit ground state 
manifolds spanned by nearest-neighbor valence bond states.  We
show that with selected perturbations such models can be driven 
into phases modeled by well understood quantum dimer models on 
the corresponding lattices. Specifically, we show that the
perturbation procedure is arbitrarily well controlled by a new
parameter which is the extent of decoration of the reference lattice.
This strategy leads to Hamiltonians that exhibit
i) $Z_2$ RVB phases in  two dimensions, ii) $U(1)$ RVB phases with
a gapless ``photon'' in three dimensions, and iii) a Cantor deconfined
region in two dimensions.
We also construct two models on the pyrochlore
lattice, one model exhibiting a $Z_2$ RVB phase and the other a 
$U(1)$ RVB phase.
\end{abstract}

\pacs{PACS numbers:
75.10-b,
75.50.Ee,
75.40.Cx,
75.40.Gb
}

\maketitle

\section{Introduction}

\begin{figure}[ht]
{\begin{center}
\includegraphics[angle=270, width=3in]{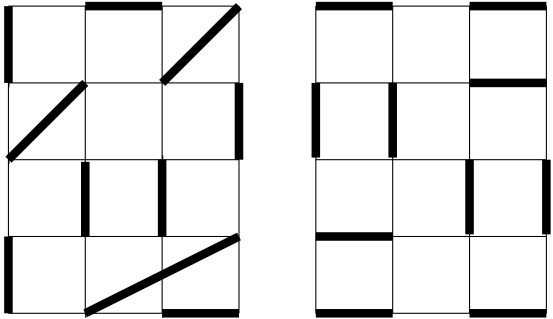}
\caption{Sample valence bond configurations.  The thick lines represent
singlet pairings.  The original formulation allowed for bonds of 
arbitrary length (left).  Considerable progress has 
been made by restricting to configurations with only nearest-neighbor valence 
bonds (right).}
\label{fig:rvb}
\end{center}}
\end{figure}

Just over thirty years ago Anderson\cite{fazekasanderson} introduced 
the resonating valence bond (RVB) state as an alternative to N\'eel ordering
in antiferromagnets with strong quantum fluctuations. In essence,
he proposed that on a sufficiently frustrated lattice an
$S=1/2$ system would exhibit a disordered state at $T=0$, which
would be captured by a wavefunction of the form
\beq
|\psi \rangle = \sum_c A_c | c \rangle
\eeq
where $| c \rangle$ is a configuration of singlet pairings of
spins or valence bonds (Fig \ref{fig:rvb}). For sufficiently short-ranged
valence bonds this describes, in contrast to the N\'eel state,
a state with short-ranged spin correlations.

\begin{figure}[ht]
{\begin{center}
\includegraphics[angle=270, width=3in]{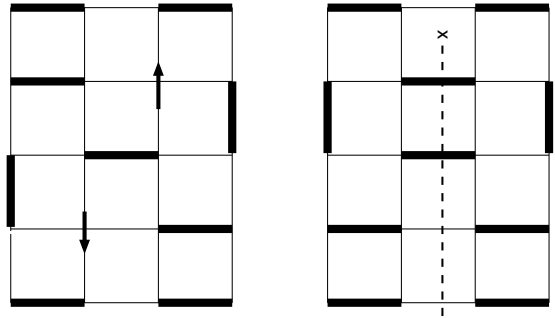}
\caption{Typical excitations of an RVB liquid.  The left figure 
depicts a pair of {\em spinons}, fractionalized excitations 
with $S=1/2$, formed by breaking a valence bond.  Spinons are unpaired 
spins which move in the RVB liquid background; the interaction 
between spinons depends on the lattice geometry.  The right figure 
depicts a {\em vison}, which is an excitation within the valence bond 
subspace.  If we consider an RVB liquid which is an equal amplitude 
superposition of nearest-neighbor valence bond states, then the vison is defined by the wavefunction 
$|\psi_{v}\rangle = \sum_c (-1)^{N_{c}} | c \rangle$ where $N_{c}$ is 
the number of bonds which cross the dashed line shown in the figure.  
Clearly this state is orthogonal to the RVB state, 
$|\psi \rangle = \sum_c |c \rangle$.  The interaction between spinons 
and visons is discussed in detail in Ref. \onlinecite{HOSToporder04}.}
\label{fig:excite}
\end{center}}
\end{figure}

The discovery of the cuprates and the suggestion that their
superconductivity could be traced to RVB physics\cite{andersonrvb} greatly
energized the elucidation of the RVB idea and by now a rather
complete understanding of its internal logic has emerged.  
In modern parlance, an RVB phase is a topological phase,
characterized by excitations with fractional quantum numbers
and a low energy gauge structure which mediates topological 
interactions among the excitations\cite{HOSToporder04}.  The 
excitations include {\em spinons}, the $S=1/2$ excitations produced by 
breaking a valence bond, as well as collective excitations within the 
valence bond manifold (see figure \ref{fig:excite}).  The sr-RVB\cite{sr-rvb} 
(short ranged RVB) with short-ranged bonds and gapped spinons will be our concern in
this paper. A post-cuprate version with longer-ranged bonds and
gapless spinons \cite{lr-RVBrevs}
has also been the subject of recent progress.\cite{hermU1}
Readers familiar with one dimensional lore will note that
the short-ranged and long-ranged RVBs generalize the physics,
respectively, of the Majumdar-Ghosh chain\cite{majghosh} and the 
Bethe chain \cite{Bethe} to higher dimensions.

The important progress that we have described has been kinematical. It has not 
directly answered the question
of realizing RVB phases for actual Hamiltonians. The original
proposal was made for the nearest-neighbor Heisenberg model on the
triangular lattice but that is now generally believed to
exhibit weak N\'eel order\cite{trineel}. To make progress on the dynamical
front, Rokhsar and Kivelson\cite{Rokhsar88} introduced the quantum dimer
model (QDM) which assumes that the low energy dynamics is
dominated by valence bond configurations of short range which
are taken to be nearest-neighbor in the versions studied to date.
Such configurations are labelled by dimer coverings of the
lattice at issue and the quantum dimer Hamiltonian acts in
a Hilbert space spanned by such coverings. The program of
studying the simplest dimer models has been rather fruitful.
It is now clear that $Z_2$ RVB phases may arise on non-bipartite 
lattices in $d\geq 2$\cite{MStrirvb}$^{,}$\cite{hkms} while bipartite lattices 
in $d>2$ give rise to $U(1)$ 
RVB phases that exhibit a gapless ``photon''\cite{msu1,pyph}. In addition a variety of
crystalline phases have been identified, most notably a
Cantor deconfined region\cite{fhmos} of interleaved commensurate and
incommensurate valence bond crystals on bipartite lattices in
$d=2$.

The next order of business then, is to find rotationally
invariant, local, spin Hamiltonians that are accurately described
by these well understood dimer models. This is the problem that
we solve in this paper thus completing a program initiated by
Chayes, Chayes and Kivelson\cite{chayeskiv}. The strategy we follow is that
of constructing Klein Hamiltonians\cite{kleinmodel} with large energy 
scales which select nearest neighbor valence bond states as their
ground states, separated by a gap from excited states. We then 
lift this degeneracy by the inclusion
of perturbations that precisely mimic the terms in the
quantum dimer models of interest. To control this procedure,
with its difficulties stemming from the non-orthogonality of
the valence bond basis, we introduce a parameter which
is the extent of decoration of the reference lattice. By suitably
tuning this parameter we are able to make our dimer model
realizations arbitrarily accurate.  An elegant feature of this limit 
is that it enables us to establish the existence of a gap about the 
nearest-neighbor valence bond manifold and to discuss the states above 
this gap in terms of ``microscopic'' spinons whose meaning will become 
clear below.  

We note that our tuning procedure occurs within the space of SU(2)
invariant Hamiltonians.  This is in contrast to approaches involving
enlarging the symmetry group to Sp($N$) or SU($N$) and studying the
large $N$ limit\cite{largeN}; in these cases, the applicability of
results to Sp(1) $\equiv$ SU(2) is not obvious.

This is also a good place to note that there is a considerable body of
work on variational\cite{eff-wavefctn} and finite-size studies of some of
the two-dimensional phases discussed in this paper, e.g.\ the early
finite-size study of a multiple-spin Hamiltonian on the
triangular lattice that adduced evidence for a topologically ordered 
phase.\cite{gregMSE} A comprehensive review
of such work is given in Ref.~\onlinecite{gregclai}. This work is
complementary to ours as it deals with somewhat simpler Hamiltonians
but is unable to access the thermodynamic limit in a controlled 
fashion. We also note that, there is a large and growing literature on more
general models with topological phases which we skip in our
focus on $S=1/2$ spin systems; this work also finds inspiration from 
the proposal that a quantum computer may be robustly created from a 
topological phase.\cite{kitaev}  Finally, we note that there are 
encouraging reports of spin liquids in experimental 
systems.\cite{coldea}$^{,}$\cite{kanoda}

In the rest of the paper we give details of our constructions.
We begin with a quick review of quantum dimer models and the
known results on their phase diagrams in Section II. In Section
III we explain our strategy with the honeycomb lattice serving
as an example; this realizes the physics of bipartite dimer
models in $d=2$. In Section IV we show how the physics
of non-bipartite dimer models in $d=2$ and bipartite and non-bipartite 
dimer models in $d>2$ may be obtained from spin models. In Section V we 
discuss two spin models on the non-bipartite pyrochlore lattice, one 
exhibiting a $Z_2$ RVB phase and the other a $U(1)$ RVB phase.  We conclude 
with a summary (Section VI) and a set of appendices that contain some technical 
material.

\section{Quantum dimer models}

For a system of spins $\vec{s}_i$ on a lattice $\Lambda$, a 
{\em (nearest-neighbor) valence bond state} 
is a product wavefunction of the form $\Psi=
\prod_{\langle ij \rangle}\psi_{ij}$ 
where $\psi_{ij}=\frac{1}{\sqrt{2}}(\psi_{i}^{\uparrow}\psi_{j}^{\downarrow}
-\psi_{i}^{\downarrow}\psi_{j}^{\uparrow})$ and the product is over 
nearest-neighbor pairs $(i,j)$.  The product is defined so each 
spin forms a singlet 
with exactly one of its neighbors.  Each valence bond state 
corresponds to a hard-core dimer covering of $\Lambda$ where a dimer 
connecting two sites corresponds to a singlet bond between the 
respective spins. Valence bond states are not orthogonal but the overlap 
between two arbitrary states is exponentially small in the length of
closed loops obtained by superposing them. This suggests that they are
linearly independent on sufficiently open lattices and indeed there
are proofs for many\cite{chayeskiv} and numerical evidence that this is so 
even on the triangular lattice\cite{matpriv}.

The identification with dimer coverings suggests that any low energy
dynamics restricted to the valence bond manifold can be represented
by a quantum dimer Hamiltonian acting on orthogonal dimer states.
The simplest such Hamiltonian on the square lattice, written down
by Rokhsar and Kivelson \cite{Rokhsar88}, has the pictorial
form
\begin{equation}
\scalebox{0.5}{\includegraphics{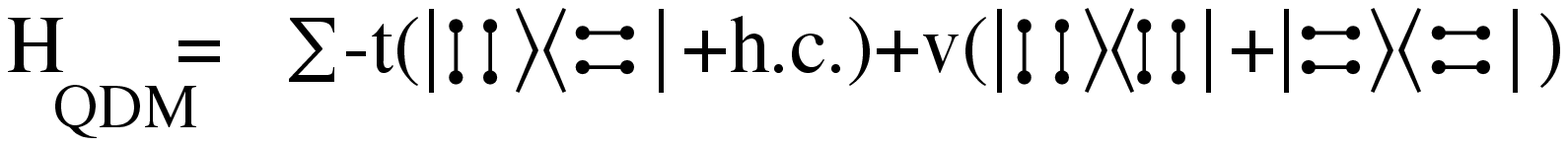}}\\
\label{eq:RKQDM}
\end{equation}
where $t$ and $v$ are positive constants and the sum is over all
possible square plaquettes. Evidently, this can be supplemented
by kinetic energy terms which act on longer loops and potential
energy terms which count more complicated dimer 
motifs.\cite{RK-footnote}

The passage from valence bonds to dimers, however, has to contend with
two complications. One is that one needs to choose a
phase convention for the valence bonds, which is subject to restrictions
on what signs one can obtain for various couplings in the dimer Hilbert
space. The other, already alluded to, is the lack of orthogonality
of the valence bond states which makes the transcription from
a spin model to a dimer model non-trivial. We will deal with both
problems later in the paper; here we merely wish to alert the
reader to their existence.

An important property of the quantum dimer Hamiltonian is the
existence of the ``Rokhsar-Kivelson point'' ({\em RK point}) $t=v$, where any  
equal amplitude superposition of all dimer coverings connected
by the operation of the kinetic energy is a ground state.
To see this, note that for every flippable plaquette, the second term gives a 
penalty $v$ while
the first term gives at most a benefit of $-t$.  Nonflippable
plaquettes are destroyed by $H_{QDM}$.  This gives a lower bound for 
the ground state energy: $E_{0}\geq$ min$\{0, N_{P}(v-t)\}$, where 
$N_{P}$ is the number of plaquettes in the lattice.  The equal amplitude state 
has energy $\langle n_{fl}\rangle(v-t)$, where $\langle n_{fl} 
\rangle$ is the average number of flippable plaquettes in the state.  At $v=t$, 
this saturates the lower bound and, since the equal amplitude state is an 
eigenstate of $H_{QDM}$ (at $v=t$), we may conclude that it is the ground state 
when $v=t$.

Thus the ground state correlations at the RK point reduce to those
of solvable classical dimer models. Additionally, the infinite
temperature static correlations of QDMs also reduce to those of
the same classical models. These features, along with the additional
one that Hamiltonians of the form (\ref{eq:RKQDM}) can
be simulated by Monte-Carlo without any sign problems,
have been crucial to making progress in determining the phase diagrams
of the quantum models. 

As a consequence of this progress we now know that: \\
i) QDMs on bipartite lattices in $d=2$ do not exhibit an RVB phase.
The equal amplitude state present at the RK point, $v/t=1$, 
does not extend into a phase.
As $v/t$ increases, the system generically passes through a sequence of 
interleaved commensurate and incommensurate crystalline phases before 
reaching the staggered valence bond solid (VBS) phase.  These intermediate 
phases, whose measure approaches unity near the RK point, turn out to 
have deconfined monomers, a phenomenon coined Cantor deconfinement\cite{fhmos}.  
As $v/t$ decreases from unity, the system passes through a plaquette 
phase\cite{mschex} 
before undergoing a first-order transition to a columnar VBS.\\
ii) QDMs on non-bipartite lattices in $d=2$ may exhibit RVB 
phases. These are $Z_2$ RVB phases captured by a 
purely topological $BF$ theory\cite{HOSToporder04}.  
For the triangular lattice, it has 
been shown\cite{MStrirvb} that for $v/t>1$, the 
system is in a staggered VBS; for 
$v/t \leq 1$, there is a deconfined RVB liquid phase.  As $v/t$ is further reduced, there are 
probably a small number of VBS phases culminating in 
the columnar state.\\
iii) QDMs on  non-bipartite lattices in $d=3$ and higher also
exhibit  a $Z_2$ RVB phase\cite{hkms}. \\
iv) QDMs on  bipartite lattices in $d=3$ and higher exhibit
a $U(1)$ RVB phase with a gapless, linearly dispersing transverse
mode, the ``photon''\cite{msu1}. \\

We now turn to the task of constructing spin models whose low
energy dynamics is precisely captured by these dimer models.
We begin, for pedagogical specificity, with the honeycomb lattice.

\section{Honeycomb lattice: Bipartite physics in $\lowercase{d}=2$}

Our strategy for realizing dimer models proceeds in three steps.
First, we construct, following Klein, a local spin Hamiltonian 
that has valence bond states as its ground states. Next we
perturb it to obtain a QDM. Finally, we decorate the lattice 
to simplify the QDM to the well studied form (\ref{eq:RKQDM}).  In the 
high decoration limit, we show the existence of a gap and give a 
description of the spectrum in terms of spinons. 

\subsection{Klein model}

The basic idea of the Klein model consists of considering a cluster of
$z$ sites (typically, a spin and its $z-1$ neighbors) and deterring, via
an energy penalty, this cluster from having
maximal total spin $S_{tot}=z/2$. If two of the spins in the
cluster form a singlet bond, this condition is
satisfied. This is why Klein Hamiltonians naturally lead to valence
bond ground states.

In particular, for a system of spins $\vec{s}_i$ on a lattice $\Lambda$, a
Klein Hamiltonian is a sum of projection
operators $\hat{P}_{\mathcal{N}(i)}$ defined as follows.  For each
site $i$, consider the neighborhood of spins $\mathcal{N}(i)$
consisting of the spin at site $i$ and its $(z_{i}-1)$ nearest
neighbors.  Let $\hat{P}_{\mathcal{N}(i)}$ be the projector onto the
highest total spin state of the cluster.  The Klein Hamiltonian is
formally given by the expression:
\begin{equation}
H_{K}=\sum_{i\in \Lambda}\hat{P}_{\mathcal{N}(i)}\ ,
\label{eq:basicklein}
\end{equation}
with total spin of cluster $\mathcal{N}(i)$ given by:
\begin{equation}
    \vec{S}_{\mathcal{N}(i)}=
\sum_{j\in\mathcal{N}(i)}\vec{s}_j
\end{equation}
We may write $\hat{P}_{\mathcal{N}(i)}$ in terms of this operator.
For example, if $z_{i}$ is even, then:
\begin{eqnarray}
    \hat{P}_{\mathcal{N}(i)}=
C_{i}\prod_{L=0}^{z_i/2-1}\Bigl[S_{\mathcal{N}(i)}^{2}-L(L+1)
\Bigr]\ .
\label{eq:kleinprojector}
\end{eqnarray}
The total spin of this cluster will take values from $0$, 
$1$,\ldots,$(z_{i}/2)-1$,$z_{i}/2$.  The factors of this product are 
operators which sequentially annihilate all but the highest spin sector.    
The form of the operator implies that larger clusters involve higher-order 
spin interactions-- but they always remain local.  

If the constants $C_{i}$ in Eq. \ref{eq:kleinprojector} are chosen to be
positive, then $H_{K}$ will have non-negative eigenvalues.  By construction,
valence bond coverings are zero-energy ground
states of $H_{K}$.

For the honeycomb lattice,
Chayes \etal\cite{chayeskiv} have already written down the explicit form of
the Klein Hamiltonian in terms of spin operators.  Their expression, up to 
unimportant overall constants, is:
\begin{eqnarray}
    H&=&\sum_{\langle
    i,j\rangle}\vec{s}_{i}\cdot\vec{s}_{j}+\frac{1}{2}\sum_{\langle\langle 
    i,j\rangle\rangle}\vec{s}_{i}\cdot\vec{s}_{j}\nonumber\\&+&\frac{2}{5}\sum_{ijkr}'
    (\vec{s}_{i}\cdot\vec{s}_{j})(\vec{s}_{k}\cdot\vec{s}_{r})
\end{eqnarray}

The first and second terms are over nearest and next nearest neighbors 
respectively.  The third term is over quartets $ijkr$ where $i$ and $j$ are 
nearest neighbors; $k$ is a neighbor of $i$ different from $j$; and 
$r$ is a neighbor of $j$ different from $i$.  A striking feature of 
this Hamiltonian is that the leading term is the usual Heisenberg 
antiferromagnet.

\subsection{Perturbations}
\label{sec:overlap}

We will now perturb the Klein Hamiltonian to obtain a QDM with
dynamics. In doing so we will use the {\it overlap expansion}
invented by Rokhsar and Kivelson\cite{Rokhsar88}.. which is predicated on the
linear independence of the valence bond states. That 
valence bond states on the honeycomb lattice are linearly
independent was proved in Ref. \onlinecite{chayeskiv}.

\begin{figure}[ht]
{\begin{center}
\includegraphics[angle=0, width=3in]{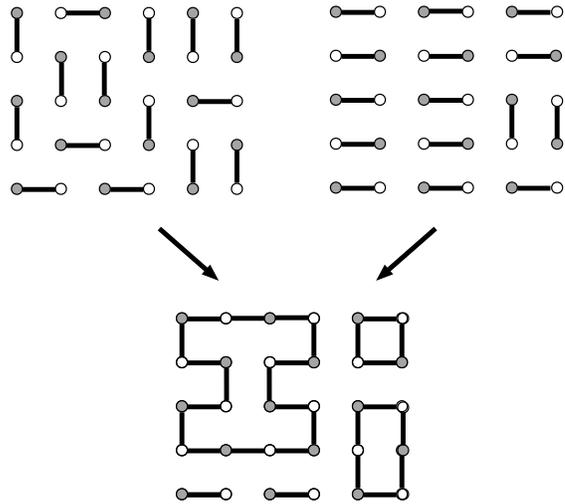}
\caption{Transition graph construction for two valence bond coverings
of the square lattice (the construction for other lattices is 
similar).  The dots are the lattice sites and the thick lines denote 
that the two sites form a singlet bond.  The singlet 
orientation may be specified, for example, by having the bonds point from 
the gray sites to the white sites.  The transition graph is formed by
overlaying the two configurations resulting in a graph (lower) containing 
double bonds and closed loops of varying (even) lengths.  In the 
above example, there are two double bonds and three loops of lengths 4, 
6, and 16.  The magnitude of the overlap between the two valence bond coverings 
is then given by $|S|=2^{3}(1/\sqrt{2})^{4+6+16}=(\frac{1}{2})^{10}=
\frac{1}{1024}$.  Thus, while the overlap between two arbitrary 
valence bond coverings is never zero, it is usually a small number.  This is 
the basis for the overlap expansion discussed in the text.}
\label{fig:transition}
\end{center}}
\end{figure}

For the purpose of obtaining the dimer kinetic energy, it
is sufficient to consider including just an additional
nearest neighbor interaction\cite{x-footnote},
\begin{equation}
    \delta H=\sum_{\langle
        i,j\rangle}\vec{s}_{i}\cdot\vec{s}_{j}
\end{equation}
To first order in degenerate perturbation theory, we may write this as an
effective operator on the valence bond states.  First, we define an orthonormal
basis set $\{|\alpha\rangle\}$ in terms of the linearly independent valence
bond states $\{|i\rangle\}$:
\begin{equation}
    |\alpha\rangle=\sum_{i}(S^{-1/2})_{\alpha,i}|i\rangle
\label{eq:ortho}
\end{equation}
Here $S_{ij}=\langle i|j\rangle$ is the overlap matrix element between
valence bond states $|i\rangle$ and $|j\rangle$.  The magnitude of the overlap
of two valence bond states may be determined by overlaying the two configurations
forming what is called the {\em transition graph}\cite{Rokhsar88}.  
The construction is described in Fig. \ref{fig:transition}.  As shown in 
Fig. \ref{fig:transition}, the transition
graph consists of double bonds, where the two states have a bond in
common, and closed loops of varying (even) lengths.  The magnitude of the
overlap $S_{ij}$ is given
by $2^{N_{l}}\prod_{i}x^{L_{i}}$ where $N_{l}$ is the number of
loops; the product is over all such loops; $L_{i}$ is the length of
loop $i$; and $x=\frac{1}{\sqrt{2}}$.  The sign of $S_{ij}$ depends on how we
choose to
orient the bonds.  By orientation, we refer to the fact that a
bond between sites 1 and 2 may be interpreted as the singlet bond
$\psi_{12}=\frac{1}{\sqrt{2}}(\psi_{1}^{\uparrow}\psi_{2}^{\downarrow}-
\psi_{1}^{\downarrow}\psi_{2}^{\uparrow})$
or as $\psi_{21}=-\psi_{12}$.  The key idea of the overlap expansion
is to treat $x$ as a small expansion parameter.

We may specify the matrix elements of our effective operator in terms of
the $\{|\alpha\rangle\}$ basis:
\begin{eqnarray}
    H_{\alpha\beta}&=&(S^{-1/2}\delta H S^{-1/2})_{\alpha\beta}\\
    &=&\sum_{ij}(S^{-1/2})_{\alpha i}\langle i|\delta H|j\rangle
    (S^{-1/2})_{j\beta}
\end{eqnarray}
If either state $|i\rangle$ or $|j\rangle$ contains the bond (12), then $\langle
i|\vec{s}_{1}\cdot\vec{s}_{2}|j\rangle=-\frac{3}{4}\langle
i|j\rangle$.  If neither $|i\rangle$ nor $|j\rangle$ contains the bond
(12), then we have a non-zero matrix element only if spins 1 and
2 are members of the same loop in the transition graph.  If that is the case, then one may show that $\langle
i|\vec{s}_{1}\cdot\vec{s}_{2}|j\rangle=(-1)^{n/2}(\mp\frac{3}{4})\langle
i|j\rangle$ where $n$ is the length of the loop for the case where
spins 1 and 2 are separated by an even (odd) number of sites.

We now specialize to the honeycomb lattice.  As sketched in the Appendix, we
may orient the bonds on the honeycomb lattice so that for any two states
differing by a (minimal) length 6 loop, the overlap $\langle i|j\rangle$
has a positive sign.  On the honeycomb lattice, the sign is a matter of
convention and we could have chosen the negative sign.  Given the
positive sign convention, we conclude that our matrix
element is given by:
\begin{equation}
   \langle i|\delta H|j\rangle=
   -\frac{3n_{d}}{4}\delta_{ij}-\frac{3}{4}(2x^{6})\hexagon_{ij}¥+O(x^{10})
\end{equation}
where $n_{d}$ is the number of bonds (half the number of sites) and
$\hexagon_{ij}$ is
a matrix that is $1$ if states $|i\rangle$ and $|j\rangle$ differ by a length 6
loop and zero otherwise.  We may also expand the overlap matrix:
\begin{eqnarray}
    S_{ij}&=&\delta_{ij}+2x^{6}\hexagon_{ij}+O(x^{10})\\
    (S^{-1/2})_{ij}&=&\delta_{ij}-x^{6}\hexagon_{ij}+O(x^{8})
\end{eqnarray}
Comparing the previous line with Eq. \ref{eq:ortho}, we see that
within the overlap expansion, each $|\alpha\rangle$ has a largest
component corresponding to a unique valence bond state.  Therefore, we will
refer to the orthogonal set $\{|\alpha\rangle\}$ as the set of {\em dimer}
states corresponding to the valence bond coverings. In writing our effective
operator, we absorb the leading term
involving the number of dimers $n_{d}$ times the unit operator into
our definition of zero energy.
\begin{eqnarray}
    H_{\alpha\beta}&=&\sum_{ij}[\delta_{\alpha i}-x^{6}\hexagon_{\alpha
    i}+O(x^{8})]\nonumber\\&\times&
    [-\frac{3}{4}(2x^{6})\hexagon_{ij}+O(x^{10})]\nonumber\\&\times&
    [\delta_{j\beta}-x^{6}\hexagon_{j\beta}+O(x^{8})]\\
    &\approx&-\frac{3}{4}(2x^{6})\hexagon_{\alpha\beta}+O(x^{10})\label{eq:QDMkin}
\end{eqnarray}
We conclude that the leading term in the overlap expansion is an
operator with nonzero matrix elements only between dimer states
differing by a minimal length 6 loop.  All of these nonzero elements
have the same value $-t=-(3/4)(2x^{6})$.  Thus we have obtained
the kinetic energy operator in the quantum dimer model (QDM).
Note that we can conclude
this only because we were able to define a bond orientation
convention such that all minimal overlaps come with the same sign.
Otherwise some off-diagonal terms would have energy $t$ and the
energy bound arguments which we gave previously to conclude that there
is an RVB state at the RK point would no longer hold.  

\begin{figure}[ht]
{\begin{center}
\includegraphics[angle=0, width=2in]{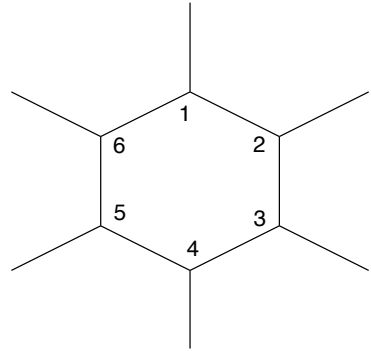}
\caption{The elementary plaquette of the honeycomb lattice.}
\label{fig:hex1}
\end{center}}
\end{figure}

To obtain the potential energy term as the leading order effect,
we need a more complicated interaction which we take to be:
\bea
    \delta H &=& J \sum_{\langle ij \rangle} \vec{s}_{i}\cdot\vec{s}_{j} + 
\nonumber\\
&&v 
    \sum_{\hexagon}\Bigl(
    (\vec{s}_{1}\cdot\vec{s}_{2})(\vec{s}_{3}\cdot\vec{s}_{4})(\vec{s}_{5}\cdot\vec{s}_{6})\nonumber\\
&&    + 
    (\vec{s}_{2}\cdot\vec{s}_{3})(\vec{s}_{4}\cdot\vec{s}_{5})(\vec{s}_{6}\cdot\vec{s}_{1})
    \Bigr)
    \label{eq:dHhex}
\eea
where the first sum is over nearest neighbors and the second sum is 
over elementary plaquettes (see Fig. \ref{fig:hex1}).

The first term of Eq.~\ref{eq:dHhex} gives the QDM kinetic energy.  
Similarly, consider the operator $s_{12}s_{34}s_{56}
=(\vec{s}_{1}\cdot\vec{s}_{2})(\vec{s}_{3}\cdot\vec{s}_{4})(\vec{s}_{5}
\cdot\vec{s}_{6})$.
If valence bond states $|i\rangle$ and $|j\rangle$ contain the bonds
(12), (34), and (56), then $\langle
i|s_{12}s_{34}s_{56}|j\rangle = (-\frac{3}{4})^{3}\langle i|j\rangle$.
If state $|i\rangle$ contains all three bonds, then the diagonal
matrix element $\langle i|s_{12}s_{34}s_{56}|i\rangle$ is
$(-\frac{3}{4})^{3}$.  If state $|i\rangle$ is missing one or more
bonds, then the diagonal matrix element is zero unless $|i\rangle$
contains the three complementary bonds (23), (45), and (61).  In this
case, the expectation of the operator equals
$(-\frac{3}{4})^{3}x^{6}$, which is higher order in the overlap
expansion.  It may be shown that off-diagonal matrix elements evaluate
to a term proportional to the overlap of the states, the
proportionality constant being of order unity.  These
results imply that the matrix element of the ring interaction between
two valence bond states is given by:
\begin{equation}
    \Bigl(-\frac{3}{4}\Bigr)^{3}vn_{fl,i}\delta_{ij}+O(vx^{6})
\end{equation}
where $n_{fl,i}$ is the number of flippable hexagonal plaquettes in 
configuration $|i\rangle$.  We may write this in terms of dimer 
states, as discussed above.  Absorbing numerical factors 
into the constants $J$ and $v$, we arrive at our effective dimer 
Hamiltonian:
\begin{eqnarray}
    H_{\alpha\beta}&=&-Jx^{6}\hexagon_{\alpha\beta}+vn_{fl,\alpha}\delta_{\alpha\beta}+O(vx^{6}+Jx^{10})\nonumber\\
    &=&-t\hexagon_{\alpha\beta}+vn_{fl,\alpha}\delta_{\alpha\beta}+O(vx^{6}+tx^{4}) \label{eq:QDMhex}
\end{eqnarray}
where $t=Jx^{6}$ and $n_{fl,\alpha}$ is the number of flippable 
plaquettes in the valence bond state corresponding to dimer state 
$\alpha$.  If $t$ and $v$ are of order unity, then the higher
order terms will be small compared to the first two terms, which act
on our dimer states (which are really spin states) in a manner
analogous to the QDM kinetic and potential energy operators on usual
dimer states.  

For the actual problem at hand, $x=(1/\sqrt{2})$ is less than $1$ but
is by no means tiny. Hence the neglect of other terms induced by
our perturbations is not obviously justified. While we do not need
them to be zero, we do need them to be weak enough perturbations
so the analysis of Ref. \onlinecite{Rokhsar88} is justified.

What we do learn from the overlap expansion
result (\ref{eq:QDMhex}) is that the non-orthogonality is a much smaller
problem on more open lattices which involve large loops. While
the honeycomb is a good candidate on this score, to put the issue
beyond doubt we now consider a decorated version of the lattice.

\subsection{Decoration scheme}

\begin{figure}[ht]
{\begin{center}
\includegraphics[angle=0, width=3in]{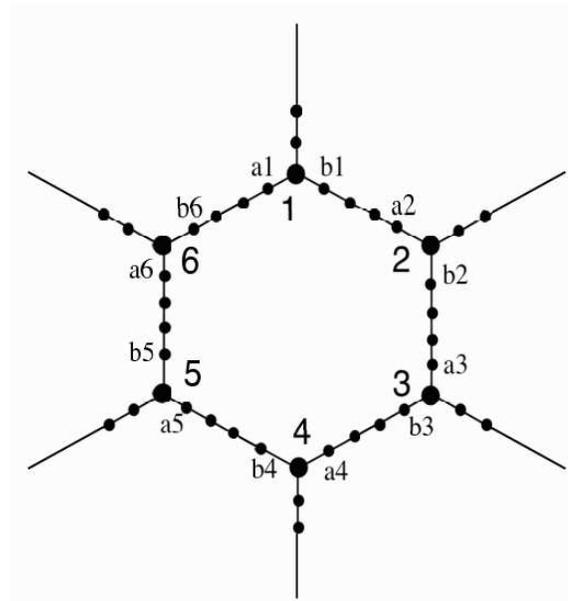}
\caption{The decorated honeycomb lattice where $N$ (an even integer)
two-fold sites are inserted between the old sites.  This drawing shows $N=4$.
The labels $a_{1}$ and $b_{1}$ designate the
first counterclockwise and first clockwise neighbor of spin 1 where
clockwise is with respect to the loop 123456.  For the undecorated
case, $a_{1}$ and $b_{1}$ are just sites 6 and 2.}
\label{fig:hex2}
\end{center}}
\end{figure}

In this section, we propose a modification to our earlier arguments
which makes the overlap expansion essentially
exact.  Consider the decorated honeycomb lattice shown in Fig.
\ref{fig:hex2} where we insert $N$ (an even integer) sites between neighboring
sites of the usual honeycomb lattice.  The dimer structure of
this lattice, including the number of dimer states, is exactly the
same as before except that having a dimer between sites 1 and 2
corresponds to a chain of $(N+2)/2$ dimers beginning at site 1 and
ending at site 2.  Not having a dimer between sites 1 and 2
corresponds to having a chain of $N/2$ dimers beginning at site
$b_{1}$ and ending at site $a_{2}$.  The Klein Hamiltonian is 
correspondingly modified by including Klein projectors for the added 
sites. 

Majumdar and Ghosh \cite{majghosh} showed that the valence bond
state is the only ground state of the Klein Hamiltonian for a
one-dimensional spin chain with an even number of spins.
Therefore, the conclusions regarding the Klein model on the honeycomb
lattice (linear independence of valence bond states, valence bond
states span the ground state manifold, etc) carry over directly to
the decorated honeycomb lattice.

While decorating does not introduce any new technical problems, there
is a significant technical advantage with respect to the overlap
expansion.  The smallest two loops on the hexagonal lattice are length
6 and 10 from which we obtained that the relative orders of the leading
and error terms in the overlap expansion were $x^{6}$ and $x^{10}$.
The smallest loops on the decorated hexagonal lattice have
lengths $6(N+1)$ and $10(N+1)$.  Repeating the previous analysis, we
will find that the leading and error terms in the overlap expansion
are $x^{6(N+1)}$ and $x^{10(N+1)}$.  The ratio of error term to
leading term has improved from $x^{4}$ to $x^{4N}$.  In the large $N$
limit, the error term is ``rigorously'' negligible but we propose that
even fairly small values of $N$ may suffice to capture the qualitative
features of the large $N$ limit.

While we have added complexity to the lattice, we do not have to
increase the order of the spin interaction.  Consider the
following as a perturbation to the decorated honeycomb lattice Klein
model:
\bea
    \delta H &=& J \sum_{\langle ij \rangle} \vec{s}_{i}\cdot\vec{s}_{j} +
\nonumber\\
&&v
    \sum_{\hexagon}\Bigl(
    (\vec{s}_{1}\cdot\vec{s}_{b_{1}})(\vec{s}_{3}\cdot\vec{s}_{b_{3}})(\vec{s}_{5}\cdot\vec{s}_{b_{5}})\nonumber\\
&&    +
    (\vec{s}_{1}\cdot\vec{s}_{a_{1}})(\vec{s}_{3}\cdot\vec{s}_{a_{3}})(\vec{s}_{5}\cdot\vec{s}_{a_{5}})
    \Bigr)
\eea
The first term is a nearest neighbor interaction over all spins while
the second term is over all elementary plaquettes, such as the one in 
Fig. \ref{fig:hex2}.  A 6 spin interaction is sufficient, even
though we have many more spins in the loop, because
having a $(1b_{1})$ bond automatically implies the other bonds in the
chain connecting 1 and 2.  Our previous analysis carries over to the
present case and we conclude:
\begin{eqnarray}
    H_{\alpha\beta}&=&-Jx^{6(N+1)}\hexagon_{\alpha\beta}+vn_{fl,\alpha}\delta_{\alpha\beta}\nonumber\\&+&O(vx^{6(N+1)}+Jx^{10(N+1)})\nonumber\\
    &=&-t\hexagon_{\alpha\beta}+vn_{fl,\alpha}\delta_{\alpha\beta}+O(vx^{6(N+1)}+tx^{4N})\nonumber\\
\end{eqnarray}
where $t=Jx^{6(N+1)}$ and otherwise the notation is the same.

Clearly, by decorating enough we can make the matrix elements beyond
the dimer model arbitarily small and thus realize the physics, including 
Cantor deconfinement, present in generic, weak perturbations of
the honeycomb QDM.

\subsubsection{Spinons}

In the highly decorated limit, one may show that nearest-neighbor 
valence bond states are separated by a finite gap from the excited 
states of the Klein model.  In this limit, we are connecting a set of 
Majumdar-Ghosh\cite{majghosh} (MG) chains into a two-dimensional 
network.  We may describe the excited states of our system in terms 
of the well studied spinon defects of the MG chains, which are widely 
believed to be gapped\cite{shassuth}$^{,}$\cite{caspers}$^{,}$\cite{sorensen}.
Here we give an outline of our argument and relegate technical 
details to Appendix \ref{app:spinongap}.

\begin{figure}[ht]
{\begin{center}
\includegraphics[angle=0, width=3in]{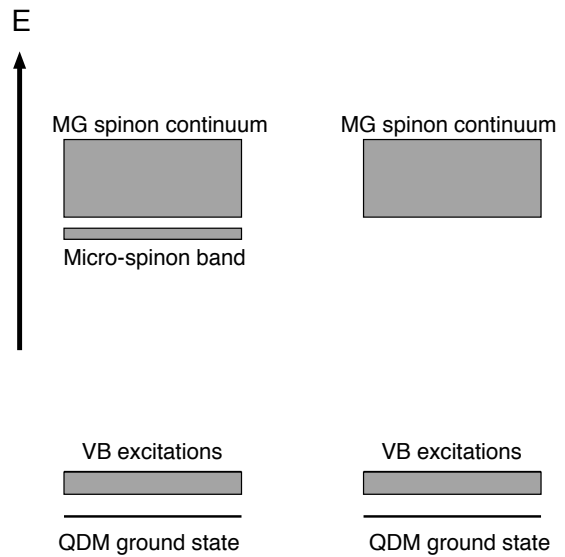}
\caption{A cartoon of the spectra for the limiting cases where the Klein ratio 
is large (left) and small (right).  In the small 
Klein ratio case (right), the lowest excited states are described by extended 
spinon wavefunctions.  These scattering states are present even in the 
large ratio case (left) but here the lowest excitations, which we call 
``microscopic'' spinons, are described by a wavefunction having peaks 
at the reference sites and decaying on the chains.  The decay rate can 
be made arbitrarily fast by tuning the Klein ratio.  The bandwidth of 
these localized states depends on the 
decoration and is zero in the infinite decoration limit.  Details are 
given in Appendix \ref{app:spinongap}.}
\label{fig:energy}
\end{center}}
\end{figure}

In Appendix \ref{app:spinongap}, we consider what happens when we put these 
chains together for different values of a tunable parameter in our model:  
the ratio of Klein scales (the coefficient $C_{i}$ in 
Eq. \ref{eq:kleinprojector}) of the Klein projectors of the decorated and 
reference sites.  For large values of this ratio, the excited states 
are represented by ``microscopic'' spinons localized on the reference 
sites.  For small values of the Klein ratio, the excited states are 
extended and may be interpreted as MG spinons scattering at the vertices.  
There is a first order transition between these limits.  In both limits, there 
is a gap between the VB manifold and the spinon states, as depicted 
in Fig. \ref{fig:energy}.

Spinons are the natural excited states (outside of the VB manifold) 
in the high decoration limit.  For an unperturbed Klein model, the VB 
manifold is degenerate so these excitations are mobile.  The next 
question is what happens when the degeneracy of the 
ground state manifold is lifted.  We argue that this has a small but 
vital effect on the spinon dynamics.  At the RK point and in liquid 
phases, we expect the spinons to be deconfined.  In the crystalline 
phases, we may consider a pair of test spinons, holding one member 
fixed and considering the quantum mechanics of the other.  If the 
wavefunction of the non-fixed spinon has spatial extent $L$, this 
would have an energy cost of order $\epsilon_c
L^{2}$ where $\epsilon_c$ is the energy cost per 
unit area of scrambling the crystalline background.  The 
$L$-dependence of the kinetic energy varies as $1/L^{2}$.  The 
implication is that while $\epsilon_c$ 
is a much smaller scale than the spinon gap, spinons moving in a 
crystalline background are confined at sufficiently long length scales.

\subsection{Square lattice}

\begin{figure}[ht]
{\begin{center}
\includegraphics[angle=0, width=2.0in]{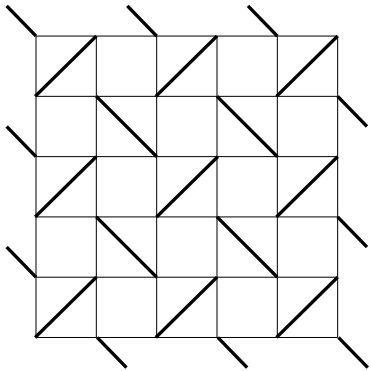}
\caption{The Klein model for the square lattice, with periodic boundary conditions, 
permits ground states which are not in manifold spanned by nearest-neighbor valence 
bond states such as this one.  Here the thin lines form the 
lattice and the thick lines denote singlet pairings.  Note that the Klein 
condition is satisfied at every lattice site.}
\label{fig:sqr1}
\end{center}}
\end{figure}

The square lattice is another two-dimensional bipartite lattice for 
which (nearest-neighbor) valence bond states are linearly 
independent\cite{chayeskiv}.  We may orient the bonds on the square 
lattice so that two states differing by a (minimal) length 4 loop, 
have positive overlap.  The problem with applying our approach to the bare 
square lattice is that the Klein model has ground states outside of the valence 
bond manifold (Fig. \ref{fig:sqr1}).  These extra states do not arise 
when we consider the decorated square lattice.  Consider perturbing 
a decorated square lattice Klein model with: 
\bea
    \delta H &=& J \sum_{\langle ij \rangle} \vec{s}_{i}\cdot\vec{s}_{j} +
\nonumber\\
&&v
    \sum_{\Box}\Bigl(
    (\vec{s}_{1}\cdot\vec{s}_{b_{1}})(\vec{s}_{3}\cdot\vec{s}_{b_{3}})+
    (\vec{s}_{1}\cdot\vec{s}_{a_{1}})(\vec{s}_{3}\cdot\vec{s}_{a_{3}})
    \Bigr)\nonumber\\
\eea
where the first term is a nearest neighbor interaction and the 
second sum is over all elementary plaquettes, the spins 1234 labelling 
the 4 sites of a square plaquette in clockwise order and, as before 
(see Fig. \ref{fig:hex2}), the labels $a_{i}$ and $b_{i}$ denoting the first 
counterclockwise and first clockwise neighbor of spin $i$.  By arguments 
similar to the honeycomb case, we obtain an effective Hamiltonian:
\begin{eqnarray}
    H_{\alpha\beta}&=&-Jx^{4(N+1)}\Box_{\alpha\beta}+vn_{fl,\alpha}\delta_{\alpha\beta}\nonumber\\&+&O(vx^{4(N+1)}+Jx^{6(N+1)})\nonumber\\
    &=&-t\Box_{\alpha\beta}+vn_{fl,\alpha}\delta_{\alpha\beta}+O(vx^{4(N+1)}+tx^{2N})\nonumber\\
\end{eqnarray}
Here $\Box_{ij}$ is a matrix that is 1 if states $|i\rangle$ and 
$|j\rangle$ differ by the (minimal) length 4 loop and zero 
otherwise.  Therefore, we realize the physics of the square lattice 
QDM.  Note that without the decoration, the error would be order 
$x^{2}=(1/2)$, as opposed to the bare honeycomb case where the error 
is order $x^{4}=(1/4)$.

\begin{figure}[ht]
{\begin{center}
\includegraphics[angle=0, width=2.0in]{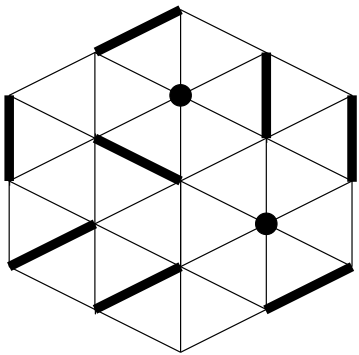}
\caption{The Klein model for the triangular lattice admits many 
nontrivial non-dimer ground states, such as this one.  The thin lines show 
the lattice and the thick lines denote singlet pairings.  The dots 
represent free spins.  Note that the Klein condition is satisfied at 
every lattice site.}
\label{fig:tri1}
\end{center}}
\end{figure}

\section{Other Valence Bond Phases in $\lowercase{d}=2$ and $\lowercase{d}=3$}

We now sketch the application of our strategy to obtain the
rest of dimer model physics, including RVB phases. The
points to be made concern the choice of lattices and phase
conventions.

\subsection{Non-bipartite lattices in $\lowercase{d}=2$}

The simplest $d=2$ non-bipartite lattice is the triangular lattice.  
Numerical evidence suggests that (nearest-neighbor) valence bond 
states are linearly independent.\cite{matpriv}  As with the square 
and honeycomb lattices, we may orient bonds so that states differing 
by a (minimal) length 4 loop, have positive overlap.\cite{MStrirvb2}   
As with the square lattice, the Klein model admits non-valence bond ground 
states, though the problem is more serious with the triangular 
lattice (see Fig. \ref{fig:tri1}).  Decoration eliminates these 
possibilities by removing the triangular nearest-neighbor 
structures.  Applying our strategy to the 
decorated triangular lattice allows us to reproduce the physics of the 
triangular lattice QDM, including its RVB phase\cite{MStrirvb}.  By 
calculations similar to Appendix \ref{app:spinongap}, one may show 
that spinon excitations are gapped.  For the triangular lattice, 
it is known that collective excitations within the valence bond 
manifold are also gapped\cite{MStrirvb}$^{,}$\cite{ivanov}, a conclusion which will remain valid for the 
decorated case.  Therefore, we have constructed a model that shows a stable, 
SU(2)-invariant RVB liquid phase.

\begin{figure}[ht]
{\begin{center}
\includegraphics[angle=270, width=3in]{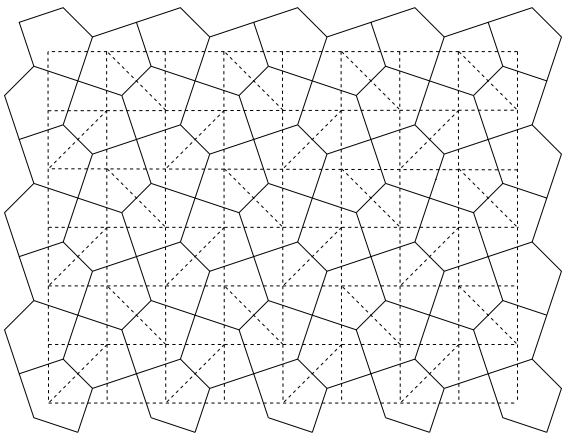}
\caption{The pentagonal (Sutherland-Shastry) lattice.  It is the dual of the Shastry-Sutherland 
lattice, which is indicated by the dashed lines.}
\label{fig:henlattice}
\end{center}}
\end{figure}

Another non-bipartite lattice is the pentagonal 
lattice\cite{fn-pentname} shown in 
Fig. \ref{fig:henlattice}.  There are currently no formal proofs for the 
pentagonal lattice regarding the issues of linear 
independence of (nearest-neighbor) valence bond states and whether the set of
these states spans the ground state space of the corresponding Klein model.
However, it was explained in Ref.~\onlinecite{chayeskiv} 
that the most important ingredients of their 
proofs for 
the honeycomb lattice are its relatively low coordination number (3); 
relatively 
large minimum loop size (6); and the absence of triangular structures in the
lattice.  The pentagonal lattice, has sites of coordination 3 and 4,
minimum loop size 8, and no triangular structures, suggesting that 
the arguments may be adapted to this lattice.  As before, it is possible to 
orient bonds so that the overlap of states differing by a (minimal) length 8 
loop always has the same sign.  While in the square, 
triangle, and honeycomb cases, the sign of the overlap is a matter of 
convention (which we chose as positive), for the pentagonal lattice, only the 
negative sign is possible.  In fact, in Appendix \ref{app:signconv}, it is shown 
that using the fermionic convention, one may always obtain the 
negative sign independent of lattice details.  Therefore, to 
generate the QDM kinetic energy, we must perturb the Klein model with a 
ferromagnetic 
interaction.  From a perturbation of the form:
\bea
    \delta H &=& -J \sum_{\langle ij \rangle} \vec{s}_{i}\cdot\vec{s}_{j} + 
\nonumber\\
&&v 
    \sum_{\octagon}\Bigl(
    (\vec{s}_{1}\cdot\vec{s}_{b_{1}})(\vec{s}_{3}\cdot\vec{s}_{b_{3}})(\vec{s}_{5}\cdot\vec{s}_{b_{5}})(\vec{s}_{7}\cdot\vec{s}_{b_{7}})\nonumber\\
&&    + 
    (\vec{s}_{1}\cdot\vec{s}_{a_{1}})(\vec{s}_{3}\cdot\vec{s}_{a_{3}})(\vec{s}_{5}\cdot\vec{s}_{a_{5}})(\vec{s}_{7}\cdot\vec{s}_{a_{7}})
    \Bigr) \label{eq:dHpent}
\eea
where the first term is over all nearest-neighbor spins and the 
second term is over elementary plaquettes ($a_{i}$ and $b_{i}$, once 
again, denoting the first counterclockwise and first clockwise 
neighbors of spin $i$), we may obtain the quantum dimer Hamiltonian 
for the decorated pentagonal lattice:
\begin{eqnarray}
    H_{\alpha\beta}&=&-t\octagon_{\alpha\beta}+vn_{fl,\alpha}\delta_{\alpha\beta}+O(vx^{8(N+1)}+tx^{2N})\nonumber\\
    \label{eq:pentQDM}
\end{eqnarray}
Here $\octagon_{ij}$ is a matrix that is 1 if states $|i\rangle$ 
and $|j\rangle$ differ by the (minimal) length 8 loop and zero 
otherwise.  Therefore, we realize the physics of the pentagonal 
lattice QDM.  We have checked that in the classical limit, the 
dimer-dimer correlations decay exponentially and in Appendix 
\ref{app:pentagonal}, we present numerical evidence that monomers are 
deconfined.  As both features also hold at the RK point, we may repeat 
the arguments described in Ref. \onlinecite{MStrirvb} for the 
triangular lattice to conclude that the pentagonal 
lattice QDM also shows an RVB liquid phase, a result which may be 
transcribed into spin language as discussed above.

\subsection{Non-bipartite lattices in $\lowercase{d}=3$}

The face-centered cubic (FCC) is a three-dimensional non-bipartite 
Bravais lattice with each site having 12 nearest-neighbors.  The undecorated 
lattice has triangular structures involving two neighboring facial 
sites and the two corners which are their common neighbors.  This will 
lead to non-valence bond ground states in the FCC Klein model.  Decoration 
eliminates the triangular structure and hence 
this type of pathology.  The 
shortest resonance loops are length 4.  In the fermion sign 
convention, these loops come with negative sign.  A perturbation 
consisting of a ferromagnetic exchange and 4-spin resonance interaction will 
reproduce the QDM results for the decorated lattice.  We expect the 
resulting model to show a $Z_{2}$ RVB phase near its RK 
point\cite{hkms}.

\subsection{Bipartite lattices in $\lowercase{d}=3$}

For the diamond lattice we pursue the same strategy as above.  The properties 
of the diamond lattice we require are the following.

The diamond lattice is bipartite, has coordination four, and the
shortest resonance loops are of length six. The Klein Hamiltonian
again has nearest-neighbour dimer coverings as ground states,
although, as for the case of the \hll, no theorem exists excluding
other ground states.  It is likely that extra states may be excluded by
decorating the lattice.  The number of dimer ground states, $n_{gs}$, is
exponentially large in the number of sites, $N$, but it is not known
exactly. Defining the ground state entropy per site as ${\cal
S}=(1/N)\ln n_{gs}$, an accurate series expansion by
Nagle\cite{nagleent} yields ${\cal S}\approx 0.265$.

We now try to mimic an RK quantum dimer model for the diamond
lattice. As for the case of the honeycomb lattice, we do this by
adding a nearest-neighbour exchange term to induce a kinetic term and
in addition, a ring term to generate a potential term. We
then expect the resulting model to exhibit, near the effective RK
point, a U(1) RVB liquid phase with algebraically decaying
correlations as well as gapless photonic gauge excitations, as
discussed in detail in Ref.~\onlinecite{msu1}.

This liquid phase will give way, upon making the potential term more
attractive, to a columnar-type solid. For an increasingly repulsive
potential, the scenario of Cantor deconfinement predicted for the
two-dimensional case is simplified. Technically, there are no relevant
lock-in terms in three dimensions so that the deconfined region simply
acquires an increasing amount of U(1) flux as $v/t$ is increased
through the RK point; finally, a staggered solid, with the maximal
amount of U(1) flux allowed by microscopic
constraints,\cite{hkms,msu1,fhmos} is reached. We cannot say whether
this will happen continuously or via a first order transition.

\section{Dynamical selection of gauge structures: 
pyrochlore lattice}

We construct a Klein model with $Z_2$ order and a Kivelson-Klein model
with $U(1)$ order which takes advantage of the bipartiteness of the
dual lattice. This nicely illustrates the dynamical selection of the
low-energy gauge structure present in topological phases.

\subsection{The Klein model}

\begin{figure}
\scalebox{0.7}{\includegraphics{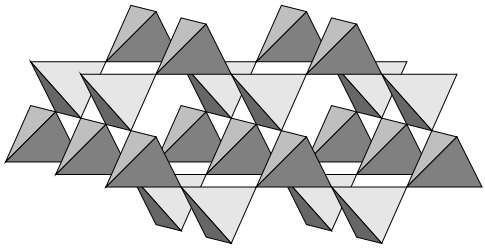}}
\caption{The pyrochlore lattice, a network of corner-sharing tetrahedra}
\label{fig:pyrochlore3d}
\end{figure}

The undecorated pyrochlore lattice (Fig.~\ref{fig:pyrochlore3d}) does not lend
itself straightforwardly as a starting point for dimer models obtained
via the Klein route because its basic building block, the
tetrahedron, supports more dimer coverings than linearly independent
singlet states.  By sufficiently decorating the lattice, the 
orthogonality problem is solved.  The shortest resonance 
loops are length 6 and the fermionic convention may be used to make 
the minimal overlaps come with negative sign.  Perturbing the 
decorated pyrochlore Klein 
model with a ferromagnetic nearest-neighbor interaction and 6-spin 
ring interaction, we obtain an effective Hamiltonian mimicking the 
pyrochlore lattice QDM, which includes a $Z_{2}$ RVB phase.   

\subsection{The Kivelson-Klein model}

A modified version of the Klein
Hamiltonian, which we will refer to as the Kivelson-Klein Hamiltonian,
may be used to produce a model displaying a $U(1)$ RVB phase.
\cite{battrug} Its Hamiltonian is of the same form as 
Eq.~\ref{eq:basicklein} but the definition of ${\cal N}(i)$ is
changed. The projection now acts not on  a site and its nearest
neighbours but instead on the four sites of a tetrahedron:
\begin{equation}
H_{KK}=\sum_{tet}\hat{P}_{tet}\ .
\label{eq:kivelsonklein}
\end{equation}
We note that the simple decoration trick described above cannot as
usefully be applied to the Kivelson-Klein model, as here increasing
the number of sites in the tetrahedron does not lead to an increase in
the number of dimers required.

\subsubsection{Ground states of the Kivelson-Klein model}

Evidently, each state in which each tetrahedron contains at least one
singlet bond is a ground state of the above Hamiltonian,
Eq.~\ref{eq:kivelsonklein}. How can this be related to dimer coverings
of the pyrochlore lattice? 

First, note that (i) the number of
tetrahedra equals twice the number of sites, $N_t=2 N$, and that (ii)
the number of hardcore dimers, $N_d\leq N/2$, as a dimer involves two
sites. From this it follows that $N_d\leq N_t$, the equality sign
holding for hardcore dimer {\em coverings}. However, the requirement
of having at least one dimer in {\em each} tetrahedron gives (iii)
$N_d\geq N_t$. We therefore see that (ii) and (iii) imply that those
hardcore dimer coverings of the pyrochlore lattice for which each
tetrahedron contains exactly one dimer are ground states of the
Kivelson-Klein Hamiltonian.

The ensemble of these states maps onto the ground states of the
six-vertex model on the diamond lattice, or equivalently onto the
ground states of the pyrochlore Ising antiferromagnet. This can be
seen as follows. First, note that the lattice of tetrahedra defined by
the pyrochlore lattice is the bipartite diamond lattice. One diamond
sublattice sits at the centre of the `up' tetrahedra, the other one at
the centres of the `down' tetrahedra. Now let us define the Ising
spins on the pyrochlore lattice as follows. For an up (down)
tetrahedra, the pair of spins at the two ends of a dimer point up
(down), and the other pair points down (up). This defines a one-to-one
mapping of dimer to Ising states; crucially, on each tetrahedron, two
spins point up and two point down, thus putting the tetrahedron into
an Ising ground state. (The mapping to the six-vertex model on the
diamond lattice proceeds by calling an up (down) spin an arrow
pointing from the centre of an up (down) tetrahedron to a down (up)
tetrahedron).

\subsubsection{Ground state correlations}

The total number of ground states (assuming that there are none in addition
to the abovementioned dimer states) gives rise to an
extensive ground state entropy well-approximated by the Pauling
entropy ${\cal S}_{Pauling}=(1/2)\ln(3/2)$.

Using the mapping to an Ising magnet, it is straightforward to
calculate the correlator between singlet bonds averaged over the
ground state manifold. To do this, 
note that each of the six dimer positions on a bond of a
given tetrahedron corresponds to an Ising ground state of that
tetrahedron. In turn, the corresponding vertex of the six-vertex model
describes a net flux, the direction of which is given as follows.
Consider a cube circumscribing the tetrahedron in question, so that
the bonds of the tetrahedron are face diagonals of the circumscribing
cube. The direction of the flux (i.e., the average direction of the
four arrows of the given vertex) now points from the centre of the
cube through the midpoint of the face of which the bond occupied by
the dimer resides.

Using the theory developed in Ref.~\onlinecite{hkms}, one can read off
that the dimer correlations are simply dipolar. Briefly, this follows
from the observation that, upon coarse graining, the smaller the
coarse-grained flux, the more microstates (prior to coarse-graining)
correspond it. Modelling this by an effective quadratic weight on the
flux configurations leads to simple magnetostatics.

For example, the connected correlator between a pair of dimers located
on the top of an up tetrahedra, separated by a vector ${\bf r}$ which
makes angle $\theta$ with the $z$ axis, is proportional to the dipolar form
$(3\cos\theta^2-1)/r^3$.

Finally, this model can again in principle be 
``Rokhsar-Kivelsonized'', i.e. by adding appropriate perturbations to 
Eq. \ref{eq:kivelsonklein}, we may generate an effective Hamiltonian 
which acts on the Kivelson-Klein ground state manifold in a manner similar 
to Eq. \ref{eq:RKQDM} on the space of dimer coverings.  We do not do this 
here for a number of reasons. Firstly, the expected
phase diagram has the same topology as that discussed for the diamond
lattice in the previous section, so no new phases are obtained. 
Secondly, the 
dimer dynamics is rather messy. The shortest
resonance loop now involves six dimers straddling a hexagonal loop of
the pyrochlore lattice -- and there are several symmetry-inequivalent
loops of this type.  (This is reminiscent of the -- exactly soluble --
kagome dimer model proposed in Ref.~\onlinecite{kagdimer}.) In
addition, for a simple perturbing nearest-neighbour exchange, the
resonance term vanishes in the leading order of the overlap expansion,
so that a more complex perturbation is needed.

\section{Discussion and outlook}

We have presented spin-1/2 Heisenberg Hamiltonians
that realize a large class of valence bond phases. In particular
they realize $Z_2$ RVB phases in $d=2$ and $d=3$, the $U(1)$
RVB phase in $d=3$ and the Cantor deconfined region in $d=2$.
These phases have previously been shown to exist in quantum
dimer models with dimers standing in for valence bonds. In
this paper we have constructed Klein models that exhibit
ground state manifolds spanned by nearest neighbor valence
bond states and then perturbed them to to realize quantum
dimer models within these manifolds.
This perturbation is done within the framework provided by
the overlap expansion, made arbitrarily 
accurate by a decoration procedure that we have introduced.

The decoration has the effect of expanding the length scale
on which the Hamiltonian acts directly. However in order
to stabilize the phases in quantum dimer models, we do
not need to go to infinite decoration -- it would be enough 
to suppress subleading terms sufficiently. In this fashion
we obtain spin models with interactions of finite 
range.\cite{cantor-fn} While large decorations are needed to
realize the simplest dimer models under analytic control,
there is little reason to doubt that the extent of decoration
could be reduced drastically, and even eliminated altogether on
some of the lattices without sacrificing the various phases
of interest. The subleading interactions will not neccessarily
uniformly tend to  subvert such phases and it is also possible
to add other terms that would stabilize them. Showing how
this can work is an obvious task for the future, as is the
construction of mathematically rigorous proofs for various
statements in this paper that are made by appealing to a small
parameter.  One promising approach is that of Ref. \onlinecite{fujimoto}, 
which appeared at the same time as our work, where the phases of the $d=2$
bipartite quantum dimer model are realized by perturbing a Klein 
model with ring exchange terms.

We emphasize that our central result has been the demonstration that 
spin liquid phases can be realized in SU(2)-invariant models.  
However, the actual models we have given involve rather complicated 
geometries and Hamiltonians without direct experimental relevance.  While it 
may be possible to engineer highly decorated lattices, a more important task 
perhaps, now that the question of 
principle is settled, is to refocus on studying much simpler Hamiltonians.
Our insistence on a specified form of the wavefunction (containing
nearest neighbor valence bonds alone) has led to fairly complicated
Hamiltonians but simpler Hamiltonians can exhibit the same phases
with more elaborate ground state wavefunctions.
Indeed, the situation for the simplest lattice under consideration in
this paper, the (undecorated) honeycomb lattice, looks quite promising.  
Previous exact
diagonalisations of a $J_1-J_2-J_3$ Heisenberg model on this
lattice\cite{fouethoney} have clearly demonstrated the existence of
the staggered VBS.  The current data appears not inconsistent with a
scenario in which the magnet leaves the staggered phase but never
reaches the fluxless plaquette phase. We are optimistic that 
it will be possible to realize the physics discussed
in this paper in such simpler models.

\section*{Acknowledgements}

We would like to thank Chris Henley for pointing out to us 
the pentagonal lattice discussed above.  We are very
grateful to Steve Kivelson for formulating the generalisation of the
Klein model on the pyrochlore lattice.  We would like to thank Matt
Hastings for a highly stimulating discussion which initiated our thoughts 
regarding the decoration procedure.  We further thank Lincoln Chayes, David Huse 
and Achim Rosch 
for useful discussions.  RM is grateful to the
Aspen Center for Physics, where parts of this work were undertaken.
This work was in part supported by the Minist\`ere de la Recherche et
des Nouvelles Technologies with an ACI grant.  SLS would like to
acknowledge support by the NSF (DMR-9978074 and 0213706) and the David
and Lucile Packard Foundation.

\appendix

\section{Sign conventions in the overlap matrix}
\label{app:signconv}

In the Rokhsar-Kivelson derivation of the dimer model, the overlap
matrix between the different dimer coverings plays a crucial
role. Whether or not the RK point corresponds to an equal-amplitude
superposition of all dimer coverings depends on the question of whether
the dimer Hamiltonian can be turned into a form where all off-diagonal
matrix elements are negative.

The leading effect of a perturbing nearest-neighbour exchange finally
gives rise to a constant (for each type of loop) multiplying the
overlap matrix (restricted to that type of loop). If the overlap
matrix can be written as a term proportional to matrix with entries only 0
or 1, the ground state is indeed given by an equal-amplitude
superposition at the RK point.

In the following, we demonstrate that a fermionic sign convention may 
be used to generate the negative sign, independent of lattice details, 
for models involving valence bond coverings of the lattice.  
We show that for the honeycomb and diamond lattices, we may 
obtain the positive sign as well.  We also present a convention for 
the Kivelson-Klein model which gives the negative sign.

\subsection{Overlaps in the fermionic convention}

A general convention for overlaps can be obtained by employing
the so-called fermionic convention, where a valence bond between sites
$a$ and $b$ is generated via operators, such as the one placing a
fermion with spin up on site $a$: $c^\dagger_{a\uparrow}$. The singlet
bond is then defined as:
\bea
|[ab]\rangle\equiv d^\dagger_{ab}|0\rangle
=\frac{1}{\sqrt{2}}[c^\dagger_{a\uparrow}c^\dagger_{b\downarrow}+
c^\dagger_{b\uparrow}c^\dagger_{a\downarrow}]|0\rangle\ .
\eea
Here $|0\rangle$ is the vacuum state with no fermions present. Note that 
\bea
d^\dagger_{ab}=d^\dagger_{ba}\ ,
\nonumber
\eea
and that these operators, being bilinear in fermions, commute unless
they have exactly one site in common. This means that for constructing
a valence-bond covering, the order in which the bonds are generated is
inconsequential.

A loop in the transition graph involving sites $h$, $i$, $j$, and $k$
will lead to the following type of expression in the overlap matrix
element calculation:
\bea
\langle 0 |
d_{ab} \cdots d_{ij}\
d^\dagger_{jk}d^\dagger_{hi} \cdots d^\dagger_{ab}
|0\rangle \nonumber\\
= 
-\frac{1}{2} \langle 0|
d_{ab} \cdots d^\dagger_{hk} \cdots d^\dagger_{ab}
|0\rangle
\eea

By induction, a loop in the transition graph involving $L$ dimers
in each configuration leads to a factor of $(-1/2)^{L-1}$,
independently of any further details of the lattice.

\subsection{Honeycomb and diamond lattices}
\label{app:hondia}

The following approach works for both honeycomb and diamond lattice,
both of which have a shortest resonance loop of length six, as in the
original benzene picture.

As these lattices are bipartite, we can orient each bond to point from
one sublattice (A) to the other (B), so that a singlet between sites
$a$ and $b$ of sublattices A and B, respectively, has the wavefunction
\bea
|(ab)\rangle\equiv
\left[
|\uparrow_a\downarrow_b\rangle-
|\downarrow_a\uparrow_b\rangle
\right]/\sqrt{2}
\ .
\eea
The two singlet coverings of the benzene loop now have wavefunctions
\bea
|1\rangle\equiv|(ab)(cd)(ef)\rangle\ \ ; \ \
|2\rangle\equiv|(bc)(de)(fa)\rangle\nonumber
\eea
from which it follows that 
\bea
\langle 1|2\rangle=+1/4
\eea 
for any hexagonal plaquette.

It is in fact also possible to choose 
\bea
\langle 1|2\rangle=-1/4
\eea 
for the honeycomb lattice. This can, for example, be achieved by
choosing any {\em fixed} hardcore dimer covering of the triangular
lattice which is dual to the honeycomb lattice under
consideration. One then multiplies each valence bond state of the
honeycomb lattice by $(-1)^{n_\times}$, where $n_\times$ is the number
of valence bonds which cross dimers of the triangular dimer
covering. This generates the desired effect.

\subsection{Other bipartite lattices}

The above construction for generating uniform overlap matrix elements
can be generalised to any bipartite lattice. By orienting the bonds
from one sublattice to the other, one always obtains an overlap which
is positive; its size is the simple product over the individual loops
involved in the transition graph.
\bea
\langle k|l\rangle>0\ ;
\eea 
indeed, the positive overlap holds true for any value of the loop
length, and therefore for an arbitrary pair of valence bond coverings
$|k\rangle$, $|l\rangle$.

\subsection{Kivelson-Klein model on pyrochlore lattice}
\label{app:KKsign}

Here we first need to establish the possible resonance loops. These
involve six dimers on a cluster of twelve sites arranged as follows.

Six of the sites sit on a hexagonal ring on the pyrochlore lattice;
each link of this hexagonal ring belongs to a different
tetrahedron. Each of these six tetrahedra contains one dimer linking a
site on the hexagonal ring with a site off the hexagonal ring. As
there is a choice of 2 such off-sites per tetrahedron, the total
number of shortest resonance loops corresponding to each hexagonal
ring equals $2^6=64$. Not all of these loops are symmetry equivalent.

These loops all involve moving six dimers. Hence, their overlap in the
Fermion convention is given by $-1/32$.

\section{Spinon gap for the decorated honeycomb lattice}
\label{app:spinongap}

A stable RVB liquid phase requires not only certain properties of the 
ground state wavefunction but also that the excitation spectrum has a 
lower bound.  In this section, we argue that the nearest neighbor
valence bond ground states are separated by a finite gap from the excited
states for the case of the decorated honeycomb lattice Klein model.

\begin{figure}[ht]
{\begin{center}
\includegraphics[angle=0, width=2.0in]{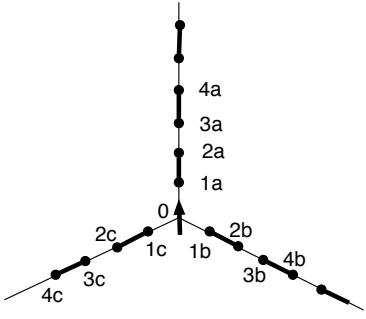}
\caption{This figure depicts a spinon at the crossing of the edges of 
a decorated honeycomb lattice.}
\label{fig:spinon0}
\end{center}}
\end{figure}

To this end observe that in the highly decorated limit we are
connecting a set of  Majumdar-Ghosh\cite{majghosh} (MG) chains into
a two-dimensional network. The excitations of the chains themselves
are well studied: these are spinon defects between the two different
dimerizations and there is considerable 
analytical\cite{shassuth}$^{,}$\cite{caspers} and
numerical\cite{sorensen} evidence that they are gapped. In putting
the chains together we need to ask if the intersections lead to
the emergence of states {\it below} the one-spinon continuum on
the chains that can fill in the gap. In the infinite decoration limit, it is 
sufficient to consider a single intersection: thus we look for
bound states localized near a site of the original honeycomb 
lattice where three MG chains would cross (see Fig. \ref{fig:spinon0}). 

We consider a variational wavefunction describing a single spin at the crossing 
of three semi-infinite MG chains.
\begin{equation}
    |\psi\rangle_{0}=|+\rangle_{0}|00\rangle_{ee}
    \label{eq:cross}
\end{equation}
where $|+\rangle_{0}$ denotes an up spin at location $0$ and 
$|00\rangle_{ee}$ denotes that ``everything else'' 
is dimerized into singlet ($00$) pairings.  We consider the action of the 
Klein Hamiltonian (\ref{eq:basicklein}) on 
this state.  Observe that $\hat{P}_{\mathcal{N}(0)}$, which involves 
only site 0 and its three neighbors, is the only projector that does not 
destroy our trial function.  A standard calculation gives:
\bea
    \hat{P}_{\mathcal{N}(0)}|\psi\rangle_{0}&=&C_{cr}\Bigl[\frac{54}{4} |\psi\rangle_{0}+
    4\Bigl(|\psi\rangle_{2a}+|\psi\rangle_{2b}+|\psi\rangle_{2c}\Bigr)+
    |\alpha\rangle\Bigr]\nonumber\\
\eea
where $|\psi\rangle_{i}=|+\rangle_{i}|00\rangle_{ee}$ describes an up 
spin at location $i$ with everything else paired into singlets and 
$C_{cr}$ sets the energy scale for violating the Klein condition at 
the cross.  We can also compute the effect of operator (\ref{eq:basicklein}) on 
the state $|\psi\rangle_{i}$, where $i$ is a location along a chain; in this case, 
only $\hat{P}_{\mathcal{N}(i)}$
contributes:
\bea
    \hat{P}_{\mathcal{N}(i)}|\psi\rangle_{i}=C_{ch}\Bigl[\frac{5}{2}|\psi\rangle_{i}
    +|\psi\rangle_{i-2}+|\psi\rangle_{i+2}+|\beta\rangle\Bigr]
\eea
Here $C_{ch}$ is the energy scale for violating the Klein condition on the 
chain.  We see that the Klein Hamiltonian acting on a spinon state produces 
the original state along with states where the spin has hopped two 
sites.  We also obtain terms which are orthogonal to all spinon 
states, designated here by the expressions $|\alpha\rangle$ and 
$|\beta\rangle$.  This motivates the bound state trial function:
\begin{equation}
    |\psi\rangle=|\psi\rangle_{0}+\sum_{n>0;i=a,b,c}y^{n}|\psi\rangle_{2n,i}
    \label{eq:boundstate}
\end{equation}
where $y$ is a variational parameter less than unity.  In calculating the 
expectation of the energy, we need to contend with the 
non-orthogonality of the spinon states:
\begin{equation}
    _{i}\langle\psi|\psi\rangle_{j}=\Bigl(-\frac{1}{2}\Bigr)^{|i\pm 
    j|/2}
\end{equation}
for sites $i$ and $j$ on the same (different) chain(s).  A tedious 
but standard calculation\cite{kumarthesis} gives the expectation value for the 
energy of the trial state (\ref{eq:boundstate}):
\begin{equation}
    E=\frac{\langle\psi|H_{K}|\psi\rangle}{\langle\psi|\psi\rangle}
\end{equation}

\bea
\frac{\langle\psi|H_{K}|\psi\rangle}{C_{cr}}&=&\frac{30+6a}{4}+3ay
\nonumber\\&+&\Bigl[\frac{18+66a}{4}+\frac{3a}{y}+12ay\Bigr]\Bigl[\frac{-y/2}{1+y/2}\Bigr]\nonumber\\
&+&\Bigl[\frac{5a}{2}+\frac{a}{y}+ay\Bigr]\Bigl[6\Bigl(\frac{-y/2}{1+y/2}\Bigr)^{2}
\nonumber\\&+&\frac{3y^{2}(1-y/2)}{(1+y/2)(1-y^{2})}\Bigr]\nonumber\\
\eea

\bea
\langle\psi|\psi\rangle&=&1+6\frac{-y/2}{1+y/2}+6\Bigl(\frac{-y/2}{1+y/2}\Bigr)^{2}
\nonumber\\&+&\frac{3y^{2}(1-y/2)}{(1+y/2)(1-y^{2})}\nonumber\\
\eea
where $a=C_{ch}/C_{cr}$. 

\begin{figure}[ht]
{\begin{center}
\includegraphics[angle=0, width=3.0in]{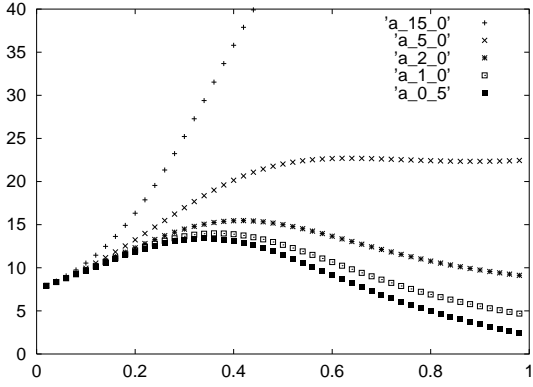}
\caption{This plot gives the energy expectation as a function of the 
variational parameter $y$ for different values of $a=C_{ch}/C_{cr}$, 
which is ratio of Klein scales; the energy is measured in units of 
$C_{cr}$.  For large values of this ratio, the minimum occurs for 
$y=0$ which corresponds to spinons localized at the crossing.  For 
small values, the minimum occurs at $y=1$, which is an extended spinon 
state.  At a value slightly greater than $a=1$, there is a first order 
``phase transition'' between these limits.  The important feature is that 
for any nonzero $a$, there is an 
energy gap between spinon states and the valence bond states, which 
have zero energy.}
\label{fig:spinongap}
\end{center}}
\end{figure}

Fig. \ref{fig:spinongap} shows a graph of this expression for several 
values of $a$.  For small values of $a$, the spinon wavefunction is an 
extended scattering state while for large values, the wavefunction is 
localized at the cross.

Our analysis has been for the infinite decoration limit.  For large, 
but finite, decoration, the extended spinon states obtained for small 
$a$ may be interpreted as MG spinons scattering at the vertices.  In
this limit, the natural extension of the localized states obtained for large $a$
will involve the wavefunction having peaks at the reference sites and 
decaying on the chains.  There will be a band of such localized states
below the scattering states.  We may estimate the bandwidth
by considering a variational wavefunction where the spinon resides 
only on the reference sites:
\begin{equation}
    |\Psi\rangle=\sum_{\vec{n}}e^{i\vec{k}\cdot\vec{n}}|\vec{n}\rangle
\end{equation}
where $|\vec{n}\rangle$ denotes a wavefunction of the form 
(\ref{eq:cross}) for the spinon at lattice site $\vec{n}$.  The 
variational energy of this trial state, to leading order in a large 
$N$ expansion, where $N$ is the number of sites inserted between 
reference sites in the decoration, is given by:
\bea
    E=C_{cr}\Bigl[\frac{54}{4}&+&16x^{2N}\Bigl(\cos k_{x}\nonumber\\&+& 
    \cos(\frac{k_{x}}{2}+\frac{\sqrt{3}k_{y}}{2})+
    \cos(\frac{k_{x}}{2}-\frac{\sqrt{3}k_{y}}{2})\Bigr)\Bigr]\nonumber\\
    \label{eq:bandwidth}
\eea
where $x=\frac{1}{\sqrt{2}}$.  We see that the band becomes more narrow 
as the decoration is increased.

\begin{figure}[ht]
{\begin{center}
\includegraphics[angle=0, width=2.0in]{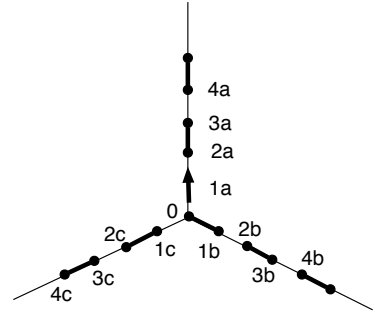}
\caption{This shows a representative of the family of states where a 
spinon is an odd distance from the origin.  In this configuration, the 
Klein operator may hop the spinon only onto the $b$-chain.  The 
configuration where the origin forms a singlet with $1c$ is 
essentially orthogonal to the given configuration for large 
decoration.}
\label{fig:spinon1}
\end{center}}
\end{figure}

In our analysis so far, we have considered states where the defects are always 
an even number of sites away from the reference sites.  There is another family of spinon 
states corresponding to the defects being located on the odd sites.  Referring to Fig. 
\ref{fig:spinon1}, the Klein operator permits the spin at 1a to hop only to 
the site 1b, which is connected by a dimer to the origin.  Therefore, in the large 
decoration limit, this is equivalent to the MG chain, which we know is 
gapped.

\begin{figure}[ht]
{\begin{center}
\includegraphics[angle=0, width=3in]{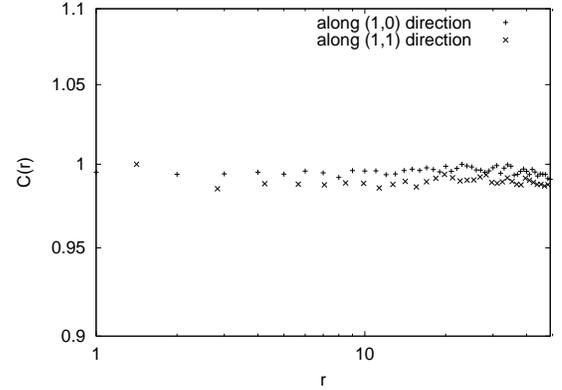}
\caption{The monomer-monomer correlation function for a 100x100 pentagonal 
lattice with periodic boundary conditions.  The two curves are cuts 
along the $\hat{x}$ and $\hat{x}+\hat{y}$ directions.  The distance 
$r$ refers to the distance between unit cells.  In computing these 
correlation functions, we take the two test monomers to be on the same 
sublattice (the pentagonal lattice is a cubic lattice with a six point 
basis).  Each data point is an average over $N=10^{6}$
data points and the noise seen in the plot is of the order of Monte 
Carlo noise $1/\sqrt{N}\sim 10^{-3}\sim 0.1\%$.}
\label{fig:monomer}
\end{center}}
\end{figure}

\section{Results for the pentagonal lattice}
\label{app:pentagonal}

At infinite temperatures, where ``infinite'' means a temperature that 
is small compared with the excitation gap of the Klein Hamiltonian 
but much larger than the energy scales of the quantum dimer 
Hamiltonian, the dimers are described classically, i.e. thermodynamic 
quantities are computed as equal-weight averages over all dimer states.  
The number of dimer states grows exponentially with lattice size.  
This number may be computed using the method of Kasteleyn. 
\cite{kasteleyn}$^{,}$\cite{fms}  The results are shown in table 
\ref{kast}.  The method also yields the entropy per site in the thermodynamic 
limit.
\begin{equation}
    S=0.168608\ldots
\end{equation}

\bigskip
\begin{table}[ht]
    \begin{tabular}{rrr}
    $L_{x}$&$L_{y}$&$n_{d}$\\
    1 &  1 & 4\\
     2 &  1 & 12\\
     3 &  1 & 28\\
     4 &  1 & 68\\
     5 &  1 & 164\\
     6 &  1 & 396\\
     1 &  2 & 12\\
     2 &  2 & 136\\
     3 &  2 & 1068\\
     4 &  2 & 9488\\
     5 &  2 & 86252\\
     6 &  2 & 798856\\
     1 &  3 & 28\\
     2 &  3 & 1068\\
     3 &  3 & 17836\\
     4 &  3 & 373412\\
     5 &  3 & 7732928\\
     6 &  3 & 160648524\\
     1 &  4 & 68\\
     2 &  4 & 9488\\
     3 &  4 & 373412\\
     4 &  4 & 21643648\\
     5 &  4 & 1235195428\\
     6 &  4 & 70937630864\\
     1 &  5 & 164\\
     2 &  5 & 86252\\
     3 &  5 & 7732928\\
     4 &  5 & 1235195428\\
     5 &  5 & 192674444864\\
     6 &  5 & 30315148743302\\
     1 &  6 & 396\\
     2 &  6 & 798856\\
     3 &  6 & 160648524\\
     4 &  6 & 70937630864\\
     5 &  6 & 30315148743302\\
 \end{tabular}

 \caption{Table of the number of dimer configurations for an 
 $L_{x}$x$L_{y}$ size pentagonal lattice.  Here $L_{x}$ and $L_{y}$ refer to 
 the underlying square lattice; the pentagonal lattice is a square 
 lattice with a 6 point basis.  We have assumed periodic boundary conditions in this 
 calculation.}
 \label{kast}
\end{table}

The striking feature of table \ref{kast} is that even fairly small 
systems have an enormous number of dimer coverings so numerical 
studies of large systems require Monte-Carlo simulations.
Fig. \ref{fig:monomer} is a Monte-Carlo 
calculation of monomer-monomer correlation functions $C(r)$ for the
pentagonal lattice using the algorithm of Sandvik\cite{sandvik}.  
$C(r)$ is defined as the number $n_{d}(r)$ of dimer coverings given a pair of 
test monomers separated by distance $r$ divided by $n_{d}(1)$. 
The simulation shows monomer deconfinement at infinite temperatures, as opposed 
to the square and honeycomb lattices which show logarithmic 
confinement.  This indicates a liquid phase at high temperatures.  The RK point 
also has deconfined monomers and the same dimer correlations as the 
high temperature phase, which strongly suggests that the RK point is part of a 
zero temperature liquid phase which connects continuously to the high 
temperature liquid.


\begin{thebibliography}{99}
\bibitem{RK-footnote}
Historically, Rokhsar and Kivelson were motivated
to consider restricting the dynamics of the nearest neighbor
antiferromagnet on the square lattice to the valence bond manifold.
This Hamiltonian is now known to have long range
order. We will consider considerably more frustrated
Hamiltonians.
\bibitem{x-footnote}
We could, in principle, select a perturbation to generate a more 
complex term at lower order in the overlap expansion described in 
section \ref{sec:overlap}.
\bibitem{fn-pentname}
In a previous paper, we have used the name pentagonal lattice for a
different tiling of the plane with irregular pentagons. As all the
pentagons in the lattice under consideration in this paper are
equivalent, we feel it is more appropriate to use the name pentagonal
lattice here.
\bibitem{cantor-fn} In the Cantor deconfined region matters are more
delicate since it involves a Cantor set of critical states. 
Nevertheless, the general argument in Ref.~\onlinecite{fhmos}
required only the weakness of a generic perturbation so that should
still go through.
\end{thebibliography}
\end{document}